\documentclass[12pt]{article}
\global\parskip 6pt
\newcommand{\bea}{\begin{eqnarray}}
\newcommand{\eea}{\end{eqnarray}}
\newcommand{\non}{\nonumber}
\newcommand{\lb}{\label}
\begin{document}
\begin{titlepage}
\hfill{hep-th/0306004}
\vspace*{.5cm}
\begin{center}
{\Large{{\bf Asymptotic Quasinormal Frequencies of d-dimensional
\\[.5ex]
Schwarzschild Black Holes}}} \\
\vspace*{1.5cm}
Danny Birmingham\footnote{Email: dannybir@strings.physics.ucsb.edu;
On leave from:
Department of Mathematical Physics, University College Dublin,
Ireland}\\
\vspace{.1cm}
{\em Department of Physics,\\
University of California, Santa Barbara,\\
Santa Barbara, CA 93106\\
USA}\\
\vspace*{1cm}
\begin{abstract}
\noindent {We determine the quasinormal frequencies for all
gravitational perturbations of the d-dimensional Schwarzschild
black hole, in the infinite damping limit. Using the potentials
for gravitational perturbations derived recently by
Ishibashi and Kodama, we show that in all cases the asymptotic real part of
the frequency is proportional to the Hawking temperature with a
coefficient of $\log 3$. Via the correspondence principle, this
leads directly to
an equally spaced entropy spectrum. We comment on the
possible implications for the spacing of eigenvalues of the Virasoro
generator in the associated near-horizon conformal algebra.}
\\
\end{abstract}
\vspace*{.25cm}
June 2003
\end{center}
\end{titlepage}

\section{Introduction}

The idea that the horizon area of black holes is quantized in equally spaced
units has attracted considerable attention 
\cite{Bekenstein}-\cite{Alekseev}.
Moreover, the possibility of a connection between the classical ringing tones (the quasinormal
frequencies) of black holes and the quantum properties of the entropy spectrum
was first observed by Bekenstein \cite{Bekenstein3}, and further developed by Hod \cite{Hod}.
In particular, Hod proposed that the real part of the quasinormal frequencies, in the
infinite damping limit, might be related via the correspondence principle
to the fundamental quanta of mass and angular momentum.
For the Schwarzschild black hole in four dimensions, the asymptotic real part
of the gravitational quasinormal frequencies is of the form $\omega = T_{H} \log 3$,
where $T_{H}$ is the Hawking temperature \cite{Motl}.
The suggestion of Hod was to identify $\hbar \omega$ with
the fundamental quantum of mass $\Delta M$.
This identification immediately leads to an area spacing of the form
$\Delta A = 4 \hbar G \log 3$.
In a separate development, Dreyer \cite{Dreyer} showed that the correspondence
principle, when applied to loop quantum gravity \cite{Ashtekar}, fixes the Immirzi parameter \cite{Immirzi}
in such a way that the Bekenstein-Hawking entropy is obtained naturally.

The proposed correspondence between quasinormal frequencies and the fundamental quantum of
mass automatically leads to an equally spaced area spectrum.
It is therefore clearly of interest to determine the universality of this approach
to black hole quantization. Although extensions to other black hole spacetimes
have been discussed \cite{Kunstatter}-\cite{Cardoso2}, the generic situation is still
far from clear. One encouraging piece of evidence comes from
an analysis of the BTZ black hole in $(2+1)$ dimensions \cite{BTZ}.
In this case, it was shown that the correspondence principle leads directly
to the correct quantum behaviour of the asymptotic Virasoro algebra \cite{BCC}.

Our aim here is to discuss the situation for the Schwarzschild black hole
in $d$ dimensions. Before considering the correspondence principle,
it is first necessary to determine the quasinormal frequencies precisely,
in the limit of infinite damping.
An elegant approach, based on analytic continuation
and computation of the monodromy of the perturbation,
was proposed by Motl and Neitzke \cite{Motl2}.
For perturbations of the $d$-dimensional Schwarzschild black hole by
a scalar field, it was found the
real part of the  asymptotic frequencies is again of the form $T_{H} \log 3$.
However, for the case of gravitational perturbations in $d > 4$, progress was impeded by
the lack of knowledge of the corresponding potentials.
Building on earlier work \cite{Ishibashi, Gibbons}, this situation has now been rectified
by the formalism of Ishibashi and Kodama \cite{Kodama}.  It has been shown that
the gravitational perturbations fall into three classes, namely scalar, vector, and
tensor perturbations, and the exact form of the potential is determined in each case.
In one application, for example, these potentials have been used to establish the
stability of the higher-dimensional Schwarzschild black hole \cite{Kodama2}.
Using the method of Motl and Neitzke, we show that the asymptotic quasinormal
frequencies of all gravitational perturbations share the $\log 3$ behaviour, in all dimensions.
This verifies the conjecture made in \cite{Motl2}, see also \cite{Kunstatter}.
By applying the correspondence principle, one is then led immediately
to an entropy spectrum with a universal $\log 3$ spacing.
We also comment on the implications of this result for the near-horizon
conformal symmetry proposed in \cite{Carlip, Carlip2, Solodukhin}.

\section{Gravitational Perturbations}

To begin, let us recall the essential features of the computation for
the case of a perturbation by a scalar field $\Phi$  satisfying
$\nabla^{2}\Phi = 0$.
The basic equation takes the form \cite{Motl2}
\bea
\left[-\left(f \frac{\partial}{\partial r}\right)^{2} + V(r) -
\omega^{2}\right] \psi(r) = 0,
\lb{scalar}
\eea
where
\bea
f = 1 - \frac{\omega_{d}M}{r^{d-3}},
\lb{f}
\eea
and $\psi = r^{(d-2)/2}\Phi$.
We have defined $\omega_{d}  = \frac{16 \pi G}{(d-2) A_{d-2}}$ in (\ref{f}),
where
$A_{d-2} = \frac{2 \pi^{(d-1)/2}}{\Gamma(\frac{d-1}{2})}$ is the volume
of the unit $(d-2)$-dimensional sphere, and we note that
$M$ has dimensions of inverse length.
The Hawking temperature of the black hole is
$T_{H} = f^{\prime}(r_{+})/4\pi = (d-3)/4 \pi r_{+}$, where
the horizon radius is defined by $r_{+}^{d-3} = \omega_{d} M$.

The potential for the scalar field perturbation is given by \cite{Motl2}
\bea
V = \frac{f}{r^{2}}\left[l(l+ d -3) +\frac{(d-2)(d-4)}{4}
+ \frac{(d-2)^{2}(\omega_{d}M)}{4 r^{d-3}}\right].
\lb{Vscalar}
\eea
The physical region of interest is $r_{+} < r < \infty$, and the quasinormal modes
are defined in terms of appropriate boundary conditions at $r = r_{+}$ and $r = \infty$.
However, the proposal of \cite{Motl2}, see also \cite{Neitzke},
is to consider an analytic continuation
to the complex $r$-plane.
It is then convenient to introduce the tortoise coordinate $z$ defined by
$dz = f^{-1}dr$,
which can be integrated to give \cite{Motl2,Gibbons}
\bea
z = r + \sum_{n=0}^{d-4}e^{2 \pi i n/(d-3)}\frac{r_{+}}{(d-3)}\log\left(1 -
\frac{r}{r_{+}}\;e^{-2\pi i n /(d-3)}\right),
\eea
where the additive constant is chosen so that $z=0$ for $r=0$.
Thus, $z$ is a multi-valued function for complex $r$.
The determination of the asymptotic quasinormal frequencies
involves a computation of the monodromy of $\psi(r)$ as one travels along a closed
contour in the complex $r$-plane.
This computation requires the ability
to match solutions in the asymptotic region and the region near
the singularity; this matching is possible precisely for the asymptotic
frequencies of interest.
The result of this local computation of the monodromy can then be compared
to the global result which follows by direct application of the
quasinormal mode boundary condition at the horizon.

For our purposes here, it is sufficient to highlight the behaviour
of $V$ in the neighbourhood of the singularity
$r=0$, namely
\bea
V \sim -\frac{(d-2)^{2}(\omega_{d}M)^{2}}{4 r^{2d-4}}.
\eea
It is straightforward to check that near $r=0$,
we have
\bea
z \sim - \frac{r^{d-2}}{(d-2)r_{+}^{d-3}}.
\lb{znear0}
\eea
Hence, the leading term in the potential
near $z=0$ is
\bea
V(z) \sim -\frac{1}{4 z^{2}},
\eea
and  Eqn. (\ref{scalar}) reduces to Bessel's equation.
By matching the solution in this region to the solution in the
asymptotic region, the monodromy can be calculated.
Comparison with the global computation of the monodromy then yields
the asymptotic quasinormal frequencies \cite{Motl2}
\bea
e^{\beta \omega} = - 3,
\lb{QNM1}
\eea
where $\beta$ is the inverse Hawking temperature.

As emphasized in \cite{Motl2}, the details of the above calculation proceed without
hindrance for the case of a potential whose behaviour near $z=0$ is
of the form
\bea V \sim \frac{j^{2} - 1}{4 z^{2}}.
\lb{Vj}
\eea
The asymptotic quasinormal frequencies in this case are given by
\bea
e^{\beta \omega} = -(1 + 2 \cos \pi j).
\lb{QNMj}
\eea

With this calculation in hand, we can now proceed to discuss the case
of gravitational perturbations.
To begin, let us consider the vector perturbation, which is
the generalization of the Regge-Wheeler equation in four dimensions \cite{Regge}.
The potential takes the form \cite{Kodama}
\bea
V_{\rm{V}} = \frac{f}{r^{2}}\left[ l(l+ d -3) +\frac{(d-2)(d-4)}{4}
- \frac{3(d-2)^{2}(\omega_{d}M)}{4 r^{d-3}}\right],
\eea
where $l\geq 2$.
We immediately notice that the leading order behaviour in the
neighbourhood of $r=0$ is given by
\bea
V_{\rm{V}} \sim \frac{3(d-2)^{2}(\omega_{d}M)^{2}}{4 r^{2d-4}}.
\eea
Using (\ref{znear0}), we see that the potential for the vector perturbation
is of the form (\ref{Vj})
with $j=2$. The asymptotic frequencies can then be simply read off from (\ref{QNMj}),
giving $e^{\beta \omega} = - 3$.

In four dimensions, the gravitational scalar perturbation is described by the
Zerilli equation \cite{Zerilli, Zerilli2}. While the form of the Zerilli potential is considerably
more complicated that the Regge-Wheeler potential, the quasinormal modes
are identical \cite{Chandrasekhar}. In higher dimensions however, we must treat
this case separately \cite{Kodama}. The potential is given by
\bea
V_{\rm{S}} = \frac{f}{r^{2}} \frac{Q}{16[c + (d-2)(d-1)x/2]^{2}},
\eea
where
\bea
Q &=& (d-2)^{4}(d-1)^{2}x^{3}\non\\
&+& (d-2)(d-1)\{4[2(d-2)^{2} - 3(d-2) + 4]c
+ (d-2)(d-4)(d-6)(d-1)\}x^{2}\non\\
&-& 12(d-2)\{(d-6)c + (d-2)(d-1)(d-4)\}cx
+
\{16c^{3} +
4(d-2)dc^{2}\},
\eea
and we have defined $c = [l(l+d-3) -(d-2)]$ and $x =
\omega_{d}M/r^{d-3}$, and again $l \geq 2$.
As in the previous case,
it is only necessary to
record the behaviour of the potential near $z=0$, which takes the form
\bea
V_{\rm{S}} \sim - \frac{(d-2)^{2} (\omega_{d}M)^{2}}{4 r^{2d-4}}.
\eea
Thus, the gravitational scalar potential is of the form (\ref{Vj})
with $j=0$. Hence,
the asymptotic frequencies again satisfy (\ref{QNM1}).
Finally, the gravitational tensor perturbations were already considered
in \cite{Motl2,Gibbons}, where it was noticed that they behave
like a scalar field perturbation.
In fact, the potential for the tensor perturbation is identical to
(\ref{Vscalar}).

In conclusion, we have shown that the asymptotic quasinormal frequencies
for gravitational perturbations of the Schwarzschild black hole
have a universal form in all dimensions, namely
\bea
\omega = \pm T_{H} \log 3 + 2 \pi i T_{H}(n + \frac{1}{2}),
\eea
as $n \rightarrow \infty$. It would be worthwhile investigating this problem
numerically, in order to verify
the results of the monodromy computation in this higher-dimensional
setting. It would also be interesting to calculate
the first order correction terms along the lines discussed
in \cite{Neitzke, Maassen}.
Incidentally, the low lying modes for perturbations by a scalar field
have been studied recently in \cite{Cardoso3, Konoplya}. 

\section{Discussion}

According to the correspondence principle \cite{Hod}, we should identify
the elementary quantum of mass $\Delta M$ with the
energy of a quantum with frequency $\omega = T_{H}\log 3$.
From the first law of thermodynamics, this leads directly
to a quantization of entropy, via the relation
\bea
\Delta S = \frac{\Delta M}{T_{H}} = \log 3.
\lb{spacing}
\eea

In \cite{Carlip,Carlip2,Solodukhin}, a conformal field theory
approach to black hole entropy in arbitrary dimensions
has been suggested. By treating the horizon
as a boundary, one finds that with a suitable
choice of boundary conditions the algebra of diffeomorphisms
in the $(r-t)$-plane near the horizon
is a Virasoro algebra. For example, in \cite{Carlip2, Park}, the
central charge and Virasoro generator
are given by
\bea
L_{0} = \frac{S}{2 \pi},\;\; \frac{c}{6} = \frac{S}{\pi},
\eea
where $S=A/4G$ is the black hole entropy.
It is then a simple matter to check that
the Cardy formula for the entropy of the conformal field theory
yields precisely the Bekenstein-Hawking entropy $S$.
This result suggests that conformal symmetry plays a key role
in understanding the microscopic properties of black holes.
Clearly, the quantization of entropy results in a corresponding spacing
of
the operator $L_{0}$, with spacing
\bea
\Delta L_{0} =
\frac{1}{2 \pi} \log 3.
\lb{L0}
\eea

In \cite{Solodukhin}, the Virasoro generator and central charge
are given by
\bea
L_{0} = \frac{1}{4 \pi^{2}q^{2}}\left(\frac{d-2}{d-3}\right)S,\;\;
\frac{c}{6} = q^{2}\left(\frac{d-3}{d-2}\right)S,
\eea
where $q$ is an arbitrary parameter. The correspondence principle
in this case gives a spacing of $L_{0}$ of the form
\bea
\Delta L_{0} =
\frac{1}{4 \pi^{2}q^{2}}\left(\frac{d-2}{d-3}\right)\log 3.
\eea
The arbitrary parameter $q$ could be fixed if one demands
integer spacing of $L_{0}$.

\noindent{\bf Acknowledgements}\\[.5ex]
D.B. is grateful to G. Horowitz for a critical reading of
the manuscript, and to the Department of Physics at UC Santa Barbara
for hospitality during the completion of this work.
This work was partially supported by Enterprise Ireland grant
IC/2002/021.

\end{document}